\documentclass[aps,prb,twocolumn,bibnotes]{revtex4-2} 

\usepackage{amsmath}
\usepackage{amssymb}
\usepackage{amsthm}
\usepackage{bbold,bm}
\usepackage[cal=boondox]{mathalfa}
\usepackage{graphicx}

\usepackage{hyperref}

\newcommand\vex[1]{\mathbf{#1}}

\def\sgn{\mathrm{sgn}}

\def\tr{\mathrm{tr}}

\def\id{\mathbb{1}} 

\usepackage{xcolor}

\begin{document} 

\title{Selective bulk-boundary correspondence in higher-order topological insulators with anticommuting mirror and chiral symmetries} 

\author{Suman Aich}
\email{aichs@iu.edu}
\affiliation{Department of Physics, Indiana University, Bloomington, Indiana 47405, USA}
\affiliation{Quantum Science and Engineering Center, Indiana University, Bloomington, Indiana 47405, USA}

\author{Babak Seradjeh}
\email{babaks@iu.edu}
\affiliation{Department of Physics, Indiana University, Bloomington, Indiana 47405, USA}
\affiliation{Quantum Science and Engineering Center, Indiana University, Bloomington, Indiana 47405, USA}

\begin{abstract}
We investigate higher-order topological insulators protected by chiral and anticommuting mirror symmetries. Using models in the BDI class, which include the prototypical topological quadrupole insulator, we show that breaking mirror symmetries that anticommute with the chiral operator leads to edge-selective bulk-boundary correspondence, with gap closings and bound states appearing only along a subset of boundaries of the same orientation and codimension. We define an edge-sensitive topological invariant that distinguishes this mechanism from previous reports of non-topological edge-selection effects.
\end{abstract}

{
\let\clearpage\relax
\maketitle
}

\section{Introduction}%
Bulk-boundary correspondence relates a topological bulk invariant of a Hermitian operator with periodic boundary conditions to the number of bound states of the same operator with open boundary conditions. Different values of the bulk invariant are separated by bulk gap closings and classify a plethora of topological phases based on global symmetries of the operator~\cite{Schnyder_2008,Kitaev_2009,Ryu_2010,Hasan_2011,Qi_2011,Chiu_2016}. Additional spatial symmetries, such as mirror reflections, can protect crystalline topological phases~\cite{Fu_2011,Slager_2012,Kruthoff_2017,Neupert_2018,Khalaf_2018}. They can also enable higher-order topological phases that have protected bound states with support on boundaries with codimension $\bar D>1$~\cite{Benalcazar_2017a,Benalcazar_2017b,Langbehn_2017,Schindler_2018}. Topological phases are then separated by gap closings in the bulk or on boundaries with codimension smaller than $\bar D$. Despite extensive work~\cite{Geier_2018,Trifunovic_2019,Calugaru_2019,Roy_2021,Khalaf_2021,Xie_2021,Yang_2024}, the full classification of higher-order topological phases and their bulk-boundary correspondence is still an open problem.

A typical property of many topological phase transitions is that they affect the boundaries of a given codimension and orientation all at once. So, for instance, if the gap along an edge of a 2d system closes and its bound state spectra change, this happens for any edge in the same direction, regardless of its relative position to the bulk. Indeed, this is what must happen when the transitions involve bulk gap closing, since the bulk spectrum cannot differentiate the relative position of its boundaries. In this work, we report topological phase transitions that are sensitive to the relative position of a boundary. In other words, the gap closing and concomitant changes to topological bound state spectra occur only selectively along certain boundaries of the same codimension and orientation. This is achieved by combining two ingredients: 1) the transitions involve gap closings along boundaries, not the bulk, hence these are higher-order topological phases; and 2) breaking crystalline symmetries that relate the boundaries with the same orientations.

We present concrete models in the BDI class that illustrate this edge-selective bulk-boundary correspondence and provide numerical results that confirm our analysis. In these models, which include the prototypical quadrupole insulator~\cite{Benalcazar_2017a,Benalcazar_2017b}, crystalline symmetries are reflections around the principal axes of the lattice, and bulk-boundary correspondence is furnished by winding numbers protected by chiral (sublattice) symmetry. Importantly, these mirror symmetries anticommute with each other and with the chiral symmetry. We establish that breaking mirror symmetries while preserving the chiral symmetry results in edge-selective gap closings and bulk-boundary correspondence furnished by an edge-selective winding number. We also discuss the breaking of diagonal mirror symmetries that commute with the chiral operator, and show that this does not lead to edge-selective gap closings. We compare our results with other instances of edge selection reported in the literature~\cite{Langbehn_2017,Geier_2018,Li_2018,Wang_2023,Kang_2024} and demonstrate that the mechanism reported here is truly topological.

The paper is organized as follows. In Sec.~\ref{sec:models} we introduce the family of models and their symmetries, and discuss patterns of mirror symmetry breaking. In Sec.~\ref{sec:topinv} we review the topological invariants proposed in the literature, examine their relationship to each other, and investigate their connection to the higher-order bulk-boundary correspondence in our models. In Sec.~\ref{sec:edgesel} we present our results on edge-selective bulk-boundary correspondence. In Sec.~\ref{sec:pert}, we investigate the effects of edge perturbations and disorder on edge-selective corner states. In Secs.~\ref{sec:nonsep} and~\ref{sec:3d}, we discuss extensions of edge-selective bulk-boundary correspondence to nonseparable and three-dimensional models. We conclude in Sec.~\ref{sec:sum} with a summary and outlook. Some details of our calculations are presented in two Appendices.

\section{Model Hamiltonians}\label{sec:models}%
\subsection{General form and symmetries}
We consider a family of 2d Bloch Hamiltonians $H(\vex k) = H_0(\vex k)+ H_\Delta(\vex k)$ over the Brillouin zone $\text{BZ} = \{\vex k = (k_1,k_2) : -\pi < k_j \leq \pi \}$ ($j=1,2$) with the tensor-product form, 
\begin{equation}\label{eq:FamHam}
H_0(\vex k) = h_1(k_1)\otimes \mathbb{1}  + c_1\otimes h_2(k_2).
\end{equation}
Here, $h_j$ are chiral Hamiltonians with chiral operators $c_j$ (for which $c_j=c_j^\dagger=c_j^{-1}$ and $c_j h_j c_j = - h_j$) with additional mirror symmetries $m_{j l}$, reflecting $k_l \to - k_l$, that commute with each other $[m_{j1},m_{j2}]=0$. Additional $\Delta$ term serves to break different symmetries, but for now we set $H_\Delta=0$. 

We also assume the commutation relations,
\begin{equation}\label{eq:CR12}
\{c_1, m_{11}\} = \{c_2, m_{22}\} = [c_1, m_{12}] = [c_2, m_{21}] =  0.
\end{equation}
Under this algebra, we can take $h_j = m_{jj} d_{je} + im_{jj}c_j d_{jo}$ where $d_{jo}$ is odd under $m_{jj}$ and both $d_{je}$ and $d_{jo}$ are otherwise even under reflections and commute with $c_j$ and $m_{jl}$. Typically, we simply have $m_{12}=m_{21}=\id$. 

These Hamiltonians are chiral under $C = c_1\otimes c_2$ and mirror-symmetric under $M_1 = m_{11}\otimes c_2 m_{21}$ and $M_2 = m_{12}\otimes m_{22}$, which all \emph{anticommute}:
\begin{align}
\{M_1,M_2\} = \{M_1,C\} = \{C,M_2\} = 0.
\end{align}
Assuming they are also time-reversal symmetric under an antiunitary operator $\Theta$ with $\Theta^2=1$, such that $\Theta H_0(-\vex k)\Theta = H_0(\vex k)$, they would also preserve particle-hole symmetry $C\Theta$ and, thus, belong to the BDI class.

For the special case $h_1(k) = h_2(k)\equiv h(k)$, the models become $C_4$ symmetric with additional anticommuting diagonal mirror symmetries $M'_1 = e^{i\frac\pi2 IP_C} C_2$ and $M'_2 = e^{i\frac\pi2 I (1-P_C)} C_1$, $\{M'_1,M'_2\}=0$, with $C_1=c_1\otimes \id$, $C_2=\id\otimes c_2$, such that $\mathsf{M}'_j(k_1,k_2)=(-1)^j(k_2,k_1)$
$
    M'_j H(\vex k) M'_j = H(\mathsf{M}'_j \vex k).
$
Here, the inversion operator $I= iM_2M_1 = i M'_2M'_1$ (note $I^\dagger= I = I^{-1}$) and the chiral projector $P_C=\frac12(1 + C)$ commute, $[I,C]=0$. We note that $M_j M'_l M_j = (-1)^{\bar j} M'_{\bar l}$ and $M'_j M_l M'_j =(-1)^j M_{\bar l}$ with $\bar j \neq j$. In this case the symmetry operator $R = e^{i\pi/4} M_1 M'_1 = e^{i\pi/4} M_2 M'_2$ is the $\pi/2$ rotation such that $R^4 = 1$ and $R^\dagger M_j R = (-1)^{j} M_{\bar j}$, and similarly $R^\dagger M'_j R = (-1)^{j} M'_{\bar j}$. Interestingly, as can be directly inspected, the diagonal mirror symmetries and the chiral operator commute, $[C,M'_j] = 0$.

\subsection{The $\pi$-flux square lattice model}
As a concrete example, we choose $h_j = d_{je}\sigma_x + d_{jo}\sigma_y$,
\begin{equation}\label{eq:ex}
    d_{je}+id_{jo} = 1+f_j + (1-f_j) e^{i n_j k_j} \equiv d_j,
\end{equation}
with $n_j\in\mathbb{Z}$ and constants $f_j\in\mathbb{R}$. 
In the following, we will assume $|f_j|<1$ for concreteness.
With this choice $|d_j(\pi/n_j)| < |d_j(0)|$.
Then, the symmetry operators are  $c_1 = c_2 = \sigma_z$, $m_{j \bar j} = \mathbb{1}$, $m_{j j} = \sigma_x$, so $C = \sigma_z\otimes\sigma_z$, $M_1 = \sigma_x \otimes \sigma_z$, $M_2 = \mathbb{1}\otimes\sigma_x$, $I = \sigma_x\otimes \sigma_y$, and $M'_1 = 
\frac12(\id\otimes \sigma_z - \sigma_z \otimes \id - \sigma_x \otimes \sigma_x + \sigma_y \otimes \sigma_y )$,
$M'_2 = 
\frac12(\id \otimes \sigma_z + \sigma_z \otimes \id + \sigma_x \otimes \sigma_x + \sigma_y \otimes \sigma_y)$. For $n_1=n_2 \equiv n$ and $f_1=f_2 \equiv f$, we have $d_1 = d_2 \equiv d$ and the model is $C_4$ symmetric. We also note that on the lattice, $M_1$ is represented by intracell hopping amplitudes in the $x_1$ direction with opposite signs, while $M_2$ is represented by intracell hopping amplitudes in the $x_2$ direction with the same sign.

The case $n_1=n_2=1$ is the $\pi$-flux square lattice model (a 2d extension of the Su-Schrieffer-Heeger model)~\cite{Seradjeh_2008b} and a topological quadrupole insulator~\cite{Benalcazar_2017a}, the first example of a higher-order topological insulator.

The low-energy theory of the $\pi$-flux square lattice model takes the Dirac form~\cite{Seradjeh_2008b,Rodriguez-Vega_2019}
\begin{equation}\label{eq:low}
    H_D = \sum_j \gamma_0 \gamma_j v_j p_j + 2|f(\vex r)| \gamma_0 e^{i\chi(\vex r)\gamma_5},
\end{equation}
where the velocities $v_j = -(1-f_j)$ in the bulk, $|f|e^{i\chi} := f_1+if_2$, and we have allowed $f_j(\vex r)$ to depend on position $\vex r=(x_1,x_2)$ near boundaries or defects, with the momentum operator $p_j = -i \partial/\partial x_j$. The Dirac matrices $\gamma_0 = \sigma_x\otimes\id$, $\gamma_1 = i\sigma_z\otimes\id$, $\gamma_2 = -i \sigma_y \otimes \sigma_y$, and $\gamma_5 = -\sigma_y \otimes \sigma_x$. Taking $\gamma_3 = i\gamma_2\gamma_1\gamma_0\gamma_5 = -i \sigma_y\otimes\sigma_z$ completes the Clifford algebra $\{ \gamma_\alpha, \gamma_\beta \} = 2g_{\alpha\beta}$ where the metric $g_{\alpha\beta} = \text{diag}(1,-1,-1,-1)$ and $\alpha, \beta \in \{0,1,2,3\}$. Then, $C = \gamma_0\gamma_3$ and $M_j = i\gamma_j\gamma_3 = -i\gamma_0\gamma_j C$. For $n_1 = n_2 = n>1$, there are $n$ Dirac points in the spectrum, so the low-energy theory is given by $n$ flavors of Dirac Hamiltonian~\eqref{eq:low}. 

\subsection{Patterns of symmetry breaking}
Now, we consider the symmetry-breaking terms in $H_\Delta$. In particular, $\Delta_l M_l$ breaks $M_{\bar l}$ and both $M'_j$, while preserving $M_l$ and $C$. Thus, combining these terms, we find that $\Delta'_1 (M_1-M_2)$ breaks both $M_j$ and $M'_2$ (preserving $M'_1$ and $C$), and $\Delta'_2 (M_1+M_2)$ breaks both $M_j$ and $M'_1$ (preserving $M'_2$ and $C$).
These terms also break the tensor-product form of $H_0$.

Other symmetry-breaking terms that break the tensor-product form are $i\delta_l M_lC_l$, which break both $M'_j$ and both $M_j$, while preserving $C$. In our concrete representation, $iM_1 C_1 = \sigma_y\otimes\sigma_z$ and $iM_2C_2 = \id\otimes\sigma_y$, so they also break time-reversal symmetry (and therefore also particle-hole symmetry). To preserve time-reversal we must multiply these terms with odd functions of $\mathbf{k}$. In this case, we can arrange to break only one of $M_j$: for example, $i \Delta_{11}C_1 M_1 \sin k_1$ preserves $M_1$ but breaks $M_2$, while $i \Delta_{12} C_1 M_1 \sin k_2$ breaks $M_1$ but preserves $M_2$.

\section{Topological invariants and bulk-boundary correspondence}\label{sec:topinv}%
\subsection{Chiral winding numbers}
In general, a chiral Hamiltonian $h$ with chiral operator $c$ can be characterized by a topological winding number $w_{\ell}[h,c]$ over a loop $\ell\subset \text{BZ}$,
\begin{equation}
w_\ell[h,c] = \frac1{2\pi i} \oint_\ell \tr[p_c\, \underline h\,\partial_k \underline h] dk,
\end{equation}
where $\underline h = 1- 2p_\text{o}$ is the flattened Hamiltonian with $p_\text{o}$ the projector to the occupied bands (i.e. with negative energies) and $p_c = \frac12 (1-c)$ is the projector to the negative chiral eigenspace~\cite{Aich_2025b}. 

\begin{figure}
    \centering
    \includegraphics[width=\linewidth]{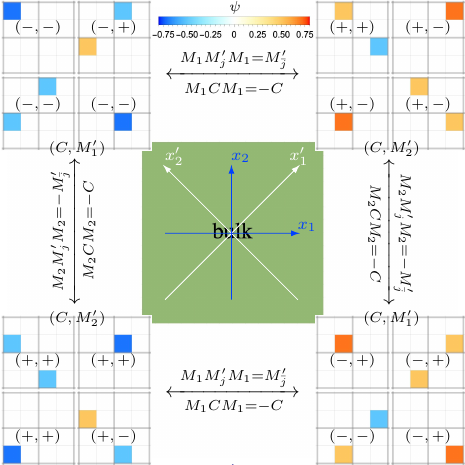}
    \caption{Eigenvalue assignments $(C,M'_j)$ of bound states in the $C_4$ symmetric model~\eqref{eq:ex} with open boundaries, $f=-0.3$ and $n_1=n_2=2$. Each square shows the numerically calculated amplitude $\psi$ of one of four mutually orthogonal bound states in a $2\times2$ mesh of unit cells at the corresponding corner. The bound states at different corners are mapped to each other by $M_j$. The coordinate systems $x_1$-$x_2$ and $x'_1$-$x'_2$ are the principal axes and diagonal directions, respectively.} \vspace{-5mm}
    \label{fig:coord}
\end{figure}

The commutation relations~\eqref{eq:CR12} ensure that $h_1$ and $h_2$ can have nontrivial winding numbers $\nu_j(k_{\bar j}) = w_{\ell_j}[h_j,c_j]$ along $\ell_j=\{\vex k : -\pi < k_j \leq \pi\}$, respectively. Thus, they constitute a stacking of 1d Hamiltonians $h_j$ along $k_{\bar j}$. 
With $\Delta=0$ (preserving the tensor-product form) the topological invariant $\nu := \nu_1\nu_2$ furnishes a bulk-boundary correspondence with $|\nu|$ zero-energy states per corner~\cite{Hayashi_2018,Hayashi_2019,Okugawa_2019}. In our concrete example Eq.~\eqref{eq:ex}, $\nu_j = \frac1{2\pi} \oint_{\ell_j} \partial_k (\log d_j) dk = n_j \theta(-f_j)$, with $\theta$ the step function. 

In $C_4$-symmetric models preserving the diagonal mirror symmetry $M'_j$, we have $[H(\vex k),M'_j]=0$ along the invariant axis of $M'_j$, i.e. $\ell'_j = \{\vex k'_{mj} := (k,(-)^jk), -\pi<k\leq\pi)\}$. Therefore, in the diagonal basis of $M'_j$ with eigenvalues $\pm1$ (where $M'_j = \sigma_z\otimes\id$) we find block-diagonal forms of the Hamiltonian $H(\vex k'_{mj}) = h'_{j+}(k) \oplus h'_{j-}(k)$ and chiral operator $C = c'_{j+}\oplus c'_{j-}$. Thus, we may define nontrivial diagonal mirror winding numbers $\nu'_{j\pm} = w_{\ell'_j}[h'_{j\pm},c'_{j\pm}]$. When both $M'_j$ are preserved, the $C_4$ symmetry enforces $\nu'_{1\pm} = \nu'_{2\pm} \equiv \nu'_{\pm}$ and $\nu'_{+} + \nu'_{-}= 0$ (see Appendix~\ref{app:winding} for proof and Ref.~\onlinecite{Aich_2025b} for related results) and the ``mirror-graded'' winding number $\nu'_{m} = (\nu'_{+}-\nu'_{-})/2 = \nu'_+$ is a stable invariant~\cite{Neupert_2018,Schindler_2018,Rodriguez-Vega_2019}. In our concrete example Eq.~\eqref{eq:ex}, we find $\nu'_m = \frac1{2\pi i} \oint_{\ell'_j} \partial_k(\log d) dk = n \theta(-f)$. 

\subsection{Bulk-boundary correspondence with $C_4$ symmetry}
Even in $C_4$-symmetric models, the question arises as to which invariants furnish the correct bulk-boundary correspondence~\cite{Okugawa_2019,Schindler_2018,Rodriguez-Vega_2019,Benalcazar_2022,Zhu_2021}. In this case, in our concrete example Eq.~\eqref{eq:ex}, $\nu = (\nu'_{m})^2$. In Fig.~\ref{fig:coord}, we visualize the number of corner bound states and their eigenvalue assignments, as confirmed in our numerical simulations~\cite{Aich_2025a}. The bulk invariants $\nu = 4$ and $\nu'_m=2$ correspond to the difference between the number of bound states with $\pm$ eigenvalues of $C$ and $M'_j$, respectively.

\begin{figure}
    \centering
    \includegraphics[width=\linewidth]{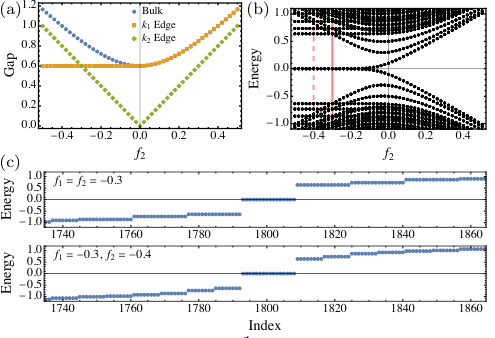}
    \caption{Bulk vs. edge gaps (a) and low-energy spectra with open boundary conditions (b),(c) for the model~\eqref{eq:ex} with $f_1=-0.3$, $n_1=n_2=2$, $\Delta=0$, and variable $f_2$.}    \vspace{-5mm}
    \label{fig:BBD0}
\end{figure}

In Fig.~\ref{fig:BBD0} we compare these invariants to the number of zero-energy states with open boundary conditions for the model with $n_1=n_2=2$~\cite{Aich_2025a}. Here, $f_1=-0.3$ is fixed. Panel (a) shows the bulk and edge gaps with periodic and cylindrical boundary conditions, respectively, and panel (b) shows the energy spectrum with open boundary conditions, as a function of $f_2$. Any open direction has $L=60$ sites. Panel (c) illustrates that for open boundary conditions, there are 16 zero-energy states, four at each corner, for $f_1=f_2=-0.3$ as well as $f_1\neq f_2 =-0.4$ marked with solid and dashed vertical lines in (b), respectively. Thus, the correct bulk-boundary correspondence is furnished by $\nu = 4$ and not the diagonal mirror-graded invariants $\nu'_{m} = 2$, even when $f_1=f_2$. Panels (a) and (b) demonstrate that the invariant $\nu=\nu_1\nu_2$ is protected by the chiral symmetry as long as the energy gap of the edge states with open boundary conditions is not closed.

\subsection{Bulk-boundary correspondence from low-energy theory}
It is well known that the low-energy Hamiltonian~\eqref{eq:low} with vortex configurations of $f(\vex r)$ supports $|n_v|$ zero-energy bound states, where the vorticity $n_v = \oint (\partial_l \chi) dl/2\pi$ is the winding of the phase $\chi$ around the vortex core~\cite{Jackiw_1981a,Hou_2007,Seradjeh_2008b}. These bound states are eigenstates of $C$ with the same eigenvalue equal to $\sgn(n_v)$. Similarly, for $n>1$ there are $n|n_v|$ stable zero-energy bound states due to $n$ flavors of Dirac Hamiltonian~\eqref{eq:low}.

There is a close relationship between corner bound states and vortex configurations in $f(\vex r)$. For an open system, even when $f(\vex r) =$ constant in the bulk, it must vary at the edges and corners. Indeed, the edges and corners of the open system correspond directly to the domain-wall and vortex configurations of $f(\vex r)$ with $n_v = \pm1$ ~\cite{Rodriguez-Vega_2019}. Therefore, the corner states arise as bound states of these vortex configurations protected by the chiral symmetry. Similar considerations have been extended to all symmetry classes of topological phases~\cite{Teo_2010}.

\begin{figure}
    \centering
    \includegraphics[width=\linewidth]{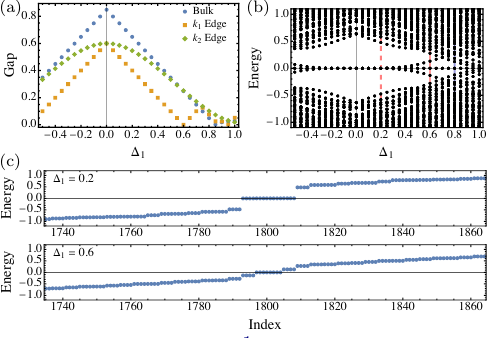}
    \caption{Bulk vs. edge gaps (a) and low-energy spectra with open boundary conditions (b),(c) for the model~\eqref{eq:ex} with $f_1=f_2=-0.3$, $n_1=n_2=2$, and $\Delta=0$ except for variable $\Delta_1$. }   \vspace{-5mm}
    \label{fig:BBD1}
\end{figure}

\section{Edge-Selective Bulk-Boundary Correspondence}\label{sec:edgesel}%
When $\ell$ is a straight line in the BZ, bulk-boundary correspondence is established by the appearance of $|w_\ell|$ zero-energy states with open boundary conditions along an edge normal to $\ell$. Since $[h,c]=0$ in the zero-energy subspace, these edge states can be chosen as eigenstates of $c$ and must all have the same eigenvalue of $c$ along a given edge. The presence of a crystalline symmetry, such as a mirror symmetry, that relates different edges can be used to determine the assignment of these eigenvalues. Considering the bulk-boundary correspondence through $\nu_j$ in our case, since $\{m_{jj},c_j\}=0$, the eigenvalues $\mathcal{c}_j$ of $c_j$ are associated with bound states $\psi_j(x_j)$ on opposite edges along the $k_{\bar j}$ direction. So, zero-energy corner states, $\psi_1(x_1)\otimes\psi_2(x_2)$, of $H_0$~\cite{Okugawa_2019} are associated with chiral eigenvalues $\mathcal{c} = \mathcal{c}_1 \mathcal{c}_2$ that alternate from one corner to the other along the edges, as in Fig.~\ref{fig:coord}. 

Interesting edge-selective gap closings become possible when the mirror symmetries are broken. For example, breaking $M_2$ and both $M'_j$ by adding $\Delta_1 M_1$, modifies $h_1 \mapsto h_1 + \mathcal{c}_2 \Delta_1 m_{11} \equiv h_1^{\mathcal{c}_2}$ where $\mathcal{c}_2 = \pm1$ is the eigenvalue of the chiral operator $c_2$. This general result can be confirmed directly in our concrete example Eq.~\eqref{eq:ex}, where $M_1 = \sigma_x\otimes\sigma_z$. Note that the tensor-product form is not preserved, and $\nu_1$ is no longer a protected topological invariant. However, since the edge states of $h_2$ (along $k_1$) are filtered according to $\mathcal{c}_2$, we might guess that increasing $|\Delta_1|$ will close the gap along only one of these edges, depending on the sign of $\Delta_1$. 

To show that this is indeed the case, we note that corner bound states of $H$ with open boundary conditions can be written as $\psi_1^{\mathcal{c}_2}(x_1)\otimes\psi_2(x_2)$ with $c_2 \psi_2 = \mathcal{c}_2\psi_2$ and $h_1^{\mathcal{c}_2} \psi_1^{\mathcal{c}_2} = 0$, assuming that $h_1^{\mathcal{c}_2}$ has zero-energy states. The latter condition is established via the bulk-boundary correspondence furnished by $\nu_1^{\mathcal{c}_2} = w_{\ell_1}[h_1^{\mathcal{c}_2},c_1]$. As we increase $|\Delta_1|$, $\nu_1^{\mathcal{c}_2}$ can change for a given value of $\mathcal{c}_2$ as the edge gap along the corresponding edge closes. Thus, we uncover an \emph{edge-selective higher-order topological transition} with bulk-boundary correspondence furnished by $\nu^{\mathcal{c}_2} = \nu^{\mathcal{c}_2}_1\nu_2$ and with zero-energy bound states at only two of the four corners. This is our main result.

In our concrete example, starting with $f_1<0$ we find a transition to the trivial phase happens for the edge with $\mathcal{c}_2 = \sgn(\Delta_1)$ at $|\Delta_1| = -2f_1$. At this value, $h_1^{\mathcal{c}_2}(\pi/n) = 0$ and the edge gap closes. Similarly, starting with $f_1>0$ an edge-selective transition to a topological phase occurs at $|\Delta_1| = 2f_1$ for the edge with $\mathcal{c}_2 = -\sgn(\Delta_1)$. 

\begin{figure}
    \centering
    \includegraphics[width=\linewidth]{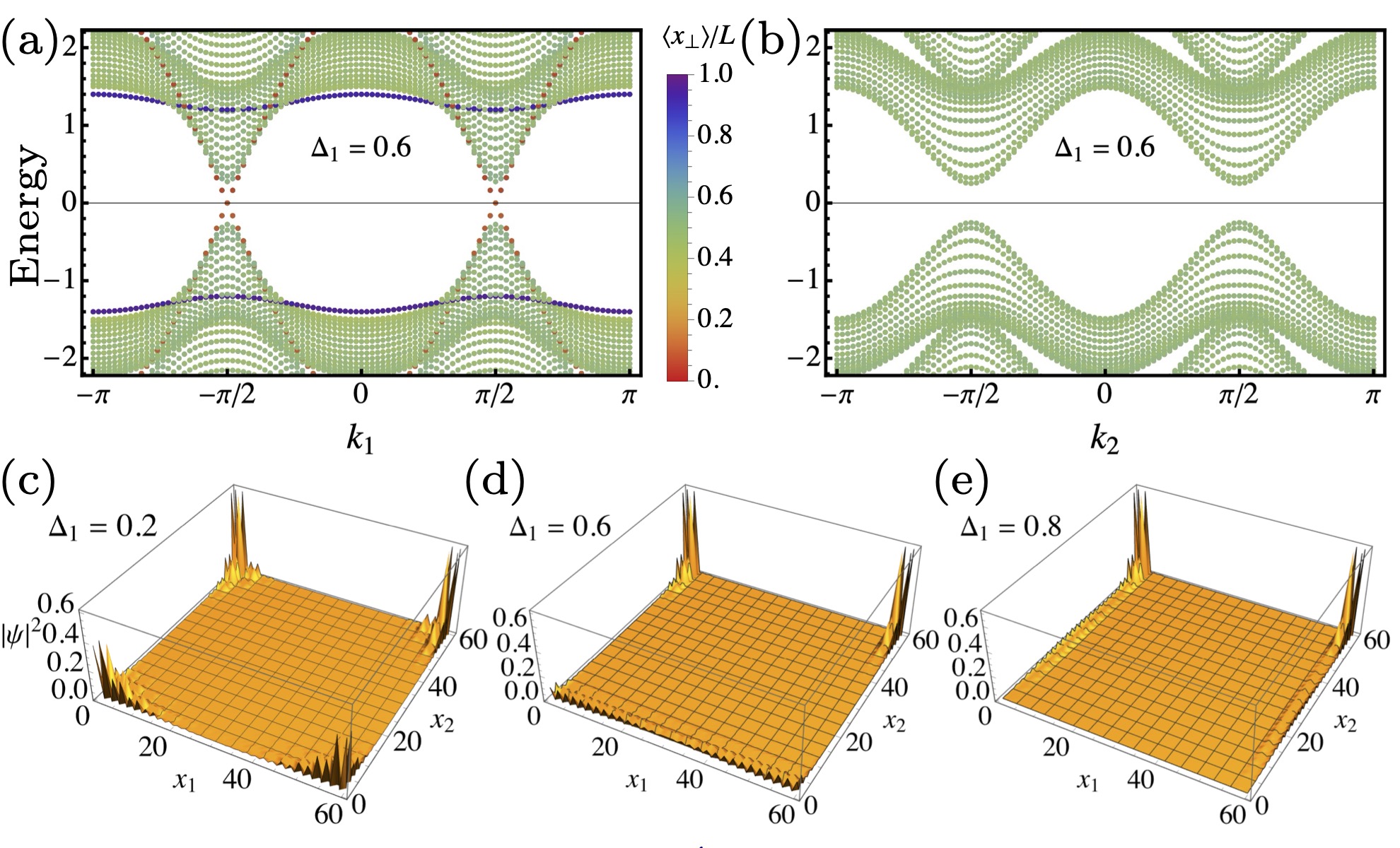}
    \caption{The spectra for $k_1$-edge (a) and $k_2$-edge (b), and corner state probability densities (c)-(e) for the model in Eq.~\eqref{eq:ex} with $f_1=f_2=-0.3$, $n_1=n_2=2$, and variable $\Delta_1$.} \vspace{-5mm}
    \label{fig:ECD1}
\end{figure}

We confirm these results in our numerical simulations for a model with $f_1=f_2=-0.3$ and $n_1=n_2=2$ as we vary $\Delta_1$. Any open direction has $L=60$ sites. In Fig.~\ref{fig:BBD1}(a), we plot the bulk and edge gaps for the model with periodic and cylindrical boundary conditions along each edge, respectively. The gap along the $k_1$ edge closes at $|\Delta_1| = -2f_1 = 0.6$, as stipulated. The gap along the $k_2$ edge and the bulk remain open below $\Delta_1\lesssim 0.85$. In Fig.~\ref{fig:BBD1}(b), we plot the energy spectra of the model with open boundary conditions. The spectra for $\Delta_1=0.2$ and $\Delta_1 = -2f_1=0.6$ are shown in Fig.~\ref{fig:BBD1}(c). Before gap closing, there are 16 zero-energy bound states; at gap closing and beyond, 8 of these states remain at zero energy while the other 8 split away to finite energies.

To examine the edge-selective nature of the gap closing, in Figs.~\ref{fig:ECD1}(a) and~\ref{fig:ECD1}(b) we plot the energy spectra at $\Delta_1=0.6$ with cylindrical boundary conditions along $k_1$ and $k_2$ edges, respectively. The color of each point on the plot shows the expectation value of position in the perpendicular direction, $x_\perp = x_2$ and $x_\perp = x_1$, respectively, for the corresponding eigenstate. This confirms that: 1) the gap remains open along the $k_2$ edge; and 2) the gap closes along the $k_1$ edge only at one edge at $x_2 = 0$ while it remains open at $x_2=L$. In Figs.~\ref{fig:ECD1}(c), (d), and (e), we plot the total probability density of the 16 states at or around zero-energy states for $\Delta_1 = 0.2$ (before gap closing), $0.6$ (at gap closing), and $0.8$ (after gap closing), respectively. This illustrates the extension and eventual merging of the bound states at corners $(0,0)$ and $(L,0)$ into the edge spectrum along the $x_2=0$ edge at gap closing. The bound states along the $x_2 = L$ edge remain localized at the corners.

One might expect that breaking both $M_j$ mirror symmetries could result in selective gap closings along two orthogonal edges, leaving a single corner state~\cite{Li_2018}. However, these selective edge gap closings are mutually exclusive. One way to see this, intuitively, is to note that once one edge gap is closed (say, along $k_1$ at $|\Delta_1| = 2|f_1|$ while keeping $\Delta_2=0$), the other edge gap can no longer close (say, by increasing $|\Delta_2|$) since there are no more corner states are left to split in pairs. Specifically, the bulk gap closes before any more edge gap closings.

To see this explicitly, we add $\Delta'_1(M_1-M_2)$ to the $C_4$ symmetric model, which breaks both $M_j$ and $M'_2$ while preserving $M'_1$. The bulk gap closes at $\Delta'_1 = -2\mathcal{m}'_1f_1$, where $\mathcal{m}'_1$ is the eigenvalue of $M'_1$, via quadratic band touchings at odd multiples of $(\pi,\pm\pi)/n$. This can be seen by noting that along the $(k,-k)$ diagonal, $[H\vert_{\ell'_1}, M'_1]=0$, and the energy eigenvalues are $\pm\sqrt2|d(k)\pm\Delta'_1|$. Along the $(k,k)$ diagonal, $[H\vert_{\ell'_{2}}, M_1-M_2]=0$, and the energy eigenvalues are $\pm\sqrt2(|d(k)|\pm|\Delta'_1|)$ (see Appendix~\ref{app:spec} for details). Therefore, for $|d(\pi/n,\pi/n)| = 2|f_1| < |\Delta'_1|<|d(0,0)| = 2$, there are $\lfloor (n+1)/2 \rfloor$ pairs of Dirac points along this diagonal. The $\lfloor n/2 \rfloor$ band touchings along the $(k,-k)$ diagonal also split into pairs of Dirac points, separated in the $(1,1)$ direction. The presence of Dirac nodes separated along the $(1,1)$ direction in the bulk spectrum yields a nonzero winding number $w_\ell[H,C]$ for $\ell \nparallel (1,1)$, corresponding to gapless spectra along any edge $\perp\ell$.

We can also study the edge-selective corner states in the low-energy theory~\eqref{eq:low}. Note that $M_1 = -i\gamma_0\gamma_1 C$, so adding a term $\Delta_1M_1$ modifies the low-energy Dirac Hamiltonian by replacing $p_1 \to p_1 - i (\Delta_1/v_1) C$. Therefore, taking a zero-energy state $\psi_0(\vex r) \propto e^{-|\vex r-\vex r_0|/R}$ of $H_D$ with chiral eigenvalue $\mathcal{c}$ and support localized in a radius of scale $R$ of the corner at $\vex r_0 = (x_0,y_0)$, the modified Hamiltonian has a corresponding zero-energy state $e^{-\mathcal{c}(\Delta_1/v_1) (x_1-x_0)}\psi_0$. This state is renormalizable for $\sgn(\mathcal{c}\Delta_1 v_1)(x_1-x_0)>0$, or for $\sgn(\mathcal{c}\Delta_1 v_1)(x_1-x_0)<0$ and $|\Delta_1/v_1| R < 1 $. Thus, while the corner bound states along an edge with $\sgn(x_1-x_0) = \sgn(\mathcal{c}\Delta_1 v_1)$ remain localized, along an edge with $\sgn(x_1-x_0) = -\sgn(\mathcal{c}\Delta_1 v_1)$ they become delocalized for $|\Delta_1| > |v_1|/R$. In the low-energy theory, we have $R = |v_1/2f_1|$ along an $x_1$ edge, hence the threshold $|v_1|/R = 2|f_1|$.

Referring to Fig.~\ref{fig:coord} for values of $\mathcal{c}$ at each corner, and recalling that $v_1 = -(1-f_1)<0$ in the topological phase, we find that the patterns of edge selection match those found above. For example, for $\Delta_1>0$, we have $\sgn(x_1-x_0) = - \sgn(\mathcal{c}) = \sgn(\mathcal{c}\Delta_1v_1)$ for corners along the $x_2=L$ edge, whereas $\sgn(x_1-x_0) = \sgn(\mathcal{c}) = - \sgn(\mathcal{c}\Delta_1v_1)$ for corners along the $x_2 = 0$ edge. Therefore increasing $\Delta_1 > 2|f_1|$ will result in delocalizing the latter, in agreement with Fig.~\ref{fig:ECD1}.

\section{Effects of Perturbation}\label{sec:pert}%
The authors of Refs.~\onlinecite{Langbehn_2017,Geier_2018} considered the selective presence of corner states in their classification of ``extrinsic'' and ``intrinsic'' higher-order topological phases, depending on lattice termination, the symmetry properties of lattice boundaries (edges and corners), and local symmetry breaking near these boundaries. An extrinsic higher-order topological phase may have corner states at generic corners that can be gapped depending on boundary decorations or terminations without bulk gap closing. An intrinsic phase, on the other hand, may host stable corner states that can only be gapped through bulk gap closings. We note that the model considered here, belonging to the BDI class, is listed in Ref.~\onlinecite{Geier_2018} as a trivial intrinsic or extrinsic higher-order insulator under inversion or a single mirror symmetry (with or without corner perturbation). It is important for our considerations that there are two anticommuting mirror symmetries in our model. The intrinsic classification under these symmetries is still trivial~\cite{luka}. One may expect that imposing the $C_4$ symmetry turns the model into an intrinsic phase, since in this case transitions are accompanied by bulk gap closings. To shed light on the nature of the edge-selective bulk-boundary correspondence, we now study the effects of boundary perturbations and disorder.

\subsection{Edge terminations}
The presence of corner states in the $\pi$-flux square lattice model at a given corner is sensitive to the lattice termination along edges forming the corner. In particular, a lattice termination through the middle of the unit cell along the edge removes or adds corner states in the topological or trivial phase, respectively. 

In fact, this is true even for the 1d Hamiltonian $h_j$ in Eq.~\eqref{eq:FamHam}, which is a first-order topologically nontrivial model in the BDI class with a $\mathbb{Z}$-valued invariant $\nu_j$, protected by the chiral symmetry $c_j$. This is so because mid-unit-cell terminations break the chiral symmetry. Instead of removing half a unit cell from the lattice, which is a singular operation that changes the Hilbert space, consider continuously increasing the local potential on the last lattice site (half of the unit cell) to a large value, say, larger than the bandwidth, which effectively removes the last site from the spectrum. This lifts the bound state on the corresponding edge to a high energy, leaving only one zero-energy bound state at the opposite edge. Such a local modification is equivalent to adding a term that projects onto the remaining half of the unit cell, proportional to $1+c_j$, which commutes with the chiral operator and breaks the chiral symmetry.

Similarly, in our models, such lattice terminations are equivalent to local modifications at the edge proportional to $1+C_1$ and $1+C_2$, depending on the direction, which break the chiral symmetry under $C = C_1 C_2$.

Some edge-selective effects have been reported in the literature recently~\cite{Wang_2023,Kang_2024}. The mechanism in these reports also relies on the fact that the eigenvalues of the chiral operator, $C$, are edge-selective. However, these works select the edge by breaking the chiral symmetry itself, e.g. by adding a term proportional to the projector $P_C=\frac12(1+C)$ to the Hamiltonian. Breaking the chiral symmetry simply shifts the energy of one set of bound states away from zero without closing any spectral gaps. As such, the reports of edge selection do not involve topological transitions. In this work, by contrast, we preserve the chiral symmetry while breaking the mirror symmetries. Our mechanism results in selective edge gap closing that involves converting extended and bound states. This is a topological phase transition.

\begin{figure}
    \centering
    \includegraphics[width=\linewidth]{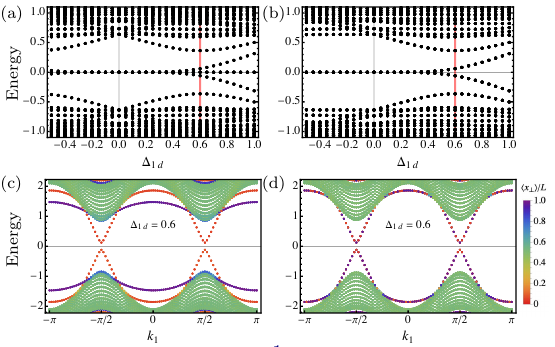}
    \caption{Spectra under edge perturbations with $\Delta_{1d}M_1$ and $\Delta_{1u}M_1$ added to two rows of unit cells at $x_2 = 0$ and $x_2 = L$, respectively, for open boundary conditions (a) and cylindrical boundary conditions in the $x_1$ direction (c) with $\Delta_{1u} = \Delta_{1d}$, and for open boundary conditions (b) and cylindrical boundary conditions in the $x_1$ direction (d) with $\Delta_{1u} = -\Delta_{1d}$. The results are shown for the $\pi$-flux square lattice  model in Eq.~\eqref{eq:ex} with $f_1=f_2=-0.3$, $n_1=n_2=2$, $L=30$ and variable $\Delta_{1d}$.} \vspace{-5mm}
    \label{fig:edgepert}
\end{figure}

\begin{figure}
    \centering
    \includegraphics[width=\linewidth]{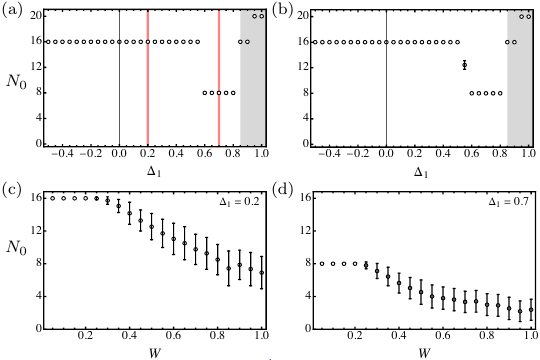}
    \caption{Number of near-zero-energy states, $N_0(\Delta_1,W)$, in the energy interval $(-\delta E, \delta E)$ as a function of uniform mirror symmetry breaking $\Delta_1$ and disorder strength $W$ for $W=0$ (a), $W=0.1$ (b), $\Delta_1 = 0.2$ (c), and $\Delta_1 = 0.7$ (d). The points show the mean and the error bars show the standard deviation calculated over 100 disorder configurations. The shaded regions in (a) and (b) mark the range for which the bulk gap remains closed. The results are shown for the $\pi$-flux square lattice  model in Eq.~\eqref{eq:ex} with $f_1=f_2=-0.3$, $n_1=n_2=2$, $L=30$ and $\delta E = 0.1$ under open boundary conditions.}
    \vspace{-5mm}
    \label{fig:disorder}
\end{figure}

\subsection{Edge and corner perturbations}
We note that breaking the protecting symmetry of a topological phase near the boundary can generically remove the boundary states. For
example, breaking time-reversal symmetry at the boundary gaps out the edge and surface states of two- and three-dimensional time-reversal invariant topological insulators without bulk gap closing.
In the following, we study perturbations that do not break the symmetries of the BDI class, in particular the chiral symmetry, but may break one or more of the protecting mirror symmetries.

We consider breaking $M_2$ mirror symmetry by adding $\Delta_1M_1$ near an edge along the $x_1$ direction. 
Because the symmetry-breaking term $\Delta_1(\vex r)$ in our analysis of the low-energy theory need only have support near the $x_1$ edge to be effective, we expect that, for appropriate sign and strength of $\Delta_1$, this will close the gap along the edge and remove the corner states along the edge.

In order to investigate the nature of edge selection, we take edge perturbations near both edges that are mapped to each other by $M_2$, with $\Delta_{1d}$ near the $x_2=0$ edge and $\Delta_{1u}$ near the $x_2=L$ edge. When $\Delta_{1u}+\Delta_{1d}\neq0$, $M_2$ is broken and we find edge selection at different values of $\Delta_{1u}$ and $\Delta_{1d}$. In particular, as shown in Fig.~\ref{fig:edgepert}(a) and (c) for maximally broken $M_2$ with $\Delta_{1u} = \Delta_{1d}$, we find gap closing along only one edge and not the other. However, when $\Delta_{1u} = - \Delta_{1d}$, $M_2$ is preserved on the lattice and, as seen in Fig.~\ref{fig:edgepert}(b) and (d), it is only possible to remove the corner states all together without edge selectivity. Note that we have chosen the parameters such that the model is $C_4$-symmetric.

On the other hand, since the corner states all have the same chiral eigenvalue, adding corner perturbations that respect the chiral symmetry cannot gap them out. This is true also for perturbations that break the mirror symmetries. Therefore, the corner states are robust against perturbations that preserve the symmetries of the BDI class and have support localized to the boundary with the same codimension. 

\begin{figure}
    \centering
    \includegraphics[width=\linewidth]{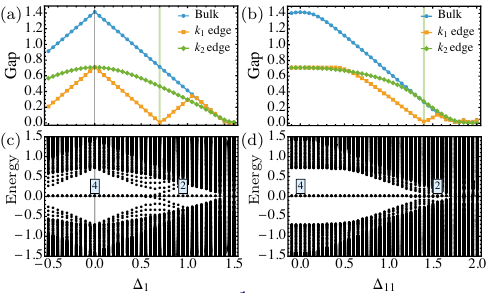}
    \caption{Bulk vs. edge spectral gaps (a),(b) and energy spectra under open boundary conditions (c),(d) for the nonseparable model in Eq.~\eqref{eq:nonsep} with $H_\Delta(\mathbf{k}) = \Delta_1M_1$ (a),(c) and $H_\Delta(\mathbf{k}) = i\Delta_{11}M_1C_1\sin k_1$ (b),(d). The boxed numbers in (c) and (d) show the number of zero-energy corner-bound states. In both cases $\bar f_1 = \bar f_2 = -0.3$, $\delta f_1 = \delta f_2 = 0.2$ and $n_1 = n_2 = 1$.} \vspace{-4mm}
    \label{fig:SEGCnonsep}
\end{figure}

\begin{figure}
    \centering
    \includegraphics[width=\linewidth]{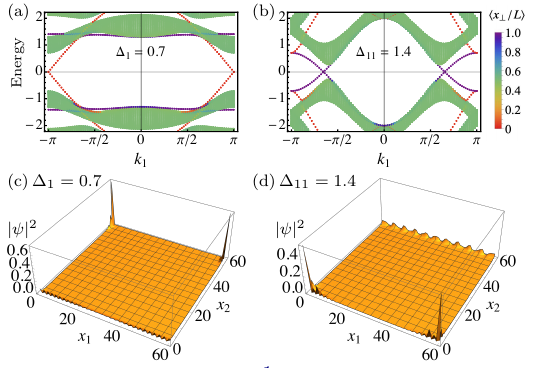}
    \caption{The $k_1$-edge spectra (a),(b) and corner state probability densities (c),(d) for parameters matching those in Fig.~\ref{fig:SEGCnonsep}.}
    \vspace{-6mm}
    \label{fig:SEGCnonsepCyl}
\end{figure}

\subsection{Disorder}
Robustness against disorder is a signature of topological phases and topological bound states that they support. Here, we examine the effects of potential disorder on the selective bulk-boundary correspondence in our model. In Fig.~\ref{fig:disorder}, we plot the number of near-zero-energy states, $N_0(\Delta_1, W)$, in the energy interval $(-\delta E, \delta E)$ in the model  with open boundary conditions, broken $M_2$ mirror symmetry by a uniform $\Delta_1M_1$ term, and random onsite potential drawn from a uniform distribution $\in [-W/2,W/2]$. As shown in Fig.~\ref{fig:disorder}(a), this quantity marks the edge-selective topological transition in the clean system at $\Delta_1 = -2f_1$. At small disorder strength in Fig.~\ref{fig:disorder}(b), we find the edge-selective topological phase transition persists. We also find the number of near-zero-energy states on each side of the phase transition, Fig.~\ref{fig:disorder}(c) and (d), is robust against disorder up to values of disorder strength comparable to the edge gap.

\section{Edge selection in nonseparable models}\label{sec:nonsep}%
We consider extensions of models in Eq.~\eqref{eq:ex} with longer-ranged and diagonal hopping elements, in which we replace $f_j \to f_j(k_j)$ and $k_j \to n_j k_j$, with $n_j\in\mathbb{Z}$ and $f_j$ real-valued even functions of their argument
\begin{equation}\label{eq:nonsep}
    d_j = 1+f_j(n_{\bar j}k_{\bar j}) + [1-f_j(n_{\bar j}k_{\bar j})] e^{i n_j k_j}.
\end{equation}
For $n_1=n_2 \equiv n$ and $f_1(k)=f_2(k) \equiv f(k)$, we have $d_1 = d_2 \equiv d$ and the model is $C_4$ symmetric. Specifically, we choose
$f_j(k) = \bar f_j + \delta f_j \cos k$.
In the following, we will assume $|f_j(k)|<1$ for simplicity. With this choice $|d_j(k_j=\pi/n_j)| < |d_j(k_j=0)|$. The case with $\delta f_j=0$ is an example of the ``separable model" discussed in the main text when $h_j$ depends only on the component of momentum $k_j$.

In Fig.~\ref{fig:SEGCnonsep}, we show the bulk and edge gaps as a function of $\Delta_1$ and $\Delta_{11}$, as well as the energy spectrum with open boundary conditions. For $\Delta_1M_1$ shown in Fig.~\ref{fig:SEGCnonsep}(a), as in the separable model, the $k_1$-edge gap closes at $|\Delta_1| \approx 0.7$, while the $k_2$-edge and bulk gaps remain open until $|\Delta_1| \approx 1.4$. For $i\Delta_{11}C_1M_1\sin k_1$, $h_1 \mapsto h_1 - i\mathcal{c}_2 \Delta_{11} m_{11}\sin k_1$ and thus by increasing $|\Delta_{11}|$, selectively closes the gap on the edge with $\mathcal c_2 = -\text{sgn}(\Delta_{11})$. Indeed, as shown in Fig.\ref{fig:SEGCnonsep}(b), the $k_1$-edge gap closes at a larger value of $|\Delta_1| \approx 1.4$, while the $k_2$-edge and bulk gaps remain open until $|\Delta_1| \approx 1.8$.

In Fig.~\ref{fig:SEGCnonsep}(c),~(d) we show the energy spectrum under open boundary conditions, with $L = 60$ sites in each direction, and the corresponding change in the number of zero-energy states across the $k_1$-edge gap closing points $|\Delta_1| \approx 0.7$ and $|\Delta_{11}| \approx 1.4$, respectively.

To examine the edge-selective nature of the $k_1$-edge gap closing, we plot the energy spectrum with cylindrical boundary conditions along $k_1$ at edge gap closings Fig.~\ref{fig:SEGCnonsepCyl}, as well as the corresponding probability densities of the near-zero-energy states. The color at each point in Fig.~\ref{fig:SEGCnonsepCyl}(a),~(b) represents the expectation value of the position of the eigenstate in the perpendicular direction, $x_\perp = x_2$. This confirms that the edge gap remains open along the $k_1$-edge closes only at the edge with $x_2 = 0$ for $\Delta_1M_1$ and $x_2 = L$ for $i\Delta_{11}C_1M_1\sin k_1$. In Fig.~\ref{fig:SEGCnonsepCyl}(c),~(d) we show the probability density of the four states around zero energy at edge gap closing. As expected from the above discussion, for $\Delta_1M_1$ the two zero-energy states localized at $x_2 = L$ remain intact at the corners, while the other two states at $x_2 = 0$ split off from zero-energy and hybridize with dispersive edge states. And the opposite happens for $i\Delta_{11}C_1M_1\sin k_1$.

\section{Edge selection in three-dimensional models}\label{sec:3d}%
We consider 3d models defined
over the Brillouin zone BZ = $\{\mathbf{k}=(k_1,k_2,k_3):-\pi\leq k_j\leq\pi\}$ $(j=1,2,3)$, with 
\begin{equation}\label{eq:H03d}
     H_0(\mathbf{k}) = h_1(\mathbf{k})\otimes\mathbb{1}\otimes\mathbb{1} + c_1\otimes h_2(\mathbf{k})\otimes \mathbb{1} + c_1\otimes c_2\otimes h_3(\mathbf{k}).
\end{equation}
The algebra of operators is similar to the 2d case: 
mirror symmetries $m_{jl}$ reflect $k_l\rightarrow -k_l$, commute with each other $[m_{jl},m_{jl'}] = 0$, and satisfy
$
    \{c_j,m_{jj}\} = [c_j,m_{jl}] = 0,
$
($j\neq l$).
These Hamiltonians are chiral under $C = c_1\otimes c_2\otimes c_3$ and mirror symmetric under $M_1 = m_{11}\otimes c_2m_{21}\otimes c_3m_{31}$, $M_2 = m_{12}\otimes m_{22}\otimes c_3m_{32}$ and $M_3 = m_{13}\otimes m_{23}\otimes m_{33}$, which mutually anticommute:
$
    \{C,M_j\} = \{M_j,M_l\} = 0,
$
($j\neq l$). Again, typically we have $m_{jl} = \mathbb{1}$ ($j\neq l$). Taking $h_j = d_{je}\sigma_x + d_{jo}\sigma_y$ as before, we have 
$
M_1 = \sigma_x \otimes \sigma_z \otimes \sigma_z,
M_2 = \id \otimes\sigma_x \otimes \sigma_z,
$
and
$M_3 = \id \otimes \id \otimes\sigma_x$.
The inversion symmetry is $I = i M_3M_2M_1$, which commutes with $M_j$ but anticommutes with $C$, and reflections in the plane $jl$ normal to $k_{l'}$ are $I_{jl} = iM_lM_j$, which commute with $M_{j'}$ and $C$. 

\begin{figure}
    \centering
    \includegraphics[width=\linewidth]{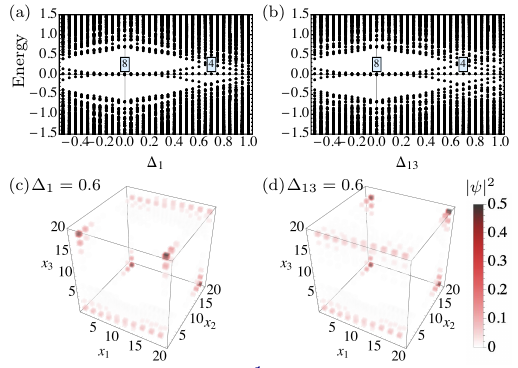}
    \caption{Open boundary energy spectra and probability densities for the separable 3d model in Eq.\eqref{eq:H03d} with (a),(c) $H_\Delta
    (\mathbf{k}) = \Delta_1 M_1$ and (b),(d) $H_\Delta
    (\mathbf{k}) = \Delta_{13} M_1C_3$. The boxed numbers in (a) and (b) show the number of zero-energy corner-bound states. In both cases $f_1 = f_2 = f_3 = -0.3$ and $n_1 = n_2 = n_3 = 1$.}
    \label{fig:SEGC3D}
\end{figure}

As a concrete example, we consider a rotationally symmetric separable model, where $d_j$ are given by Eq.~\eqref{eq:ex}, with $f_j=f$. For $f<0$ and $H_\Delta = 0$, this model is a third-order topological insulator with $n^3$ zero-energy states localized at each corner corresponding to the bulk  invariant $\nu = \nu_1\nu_2\nu_3$ where $\nu_j = n$ is the winding number of $h_j$.

There are two classes of mirror-symmetry breaking terms that break either two and one of the mirror symmetries, respectively, while preserving the chiral symmetry: $H_\Delta(\mathbf{k}) = \Delta_j M_j$ breaks the mirror symmetries $M_{j'}$ ($j'\neq j$), whereas $H_\Delta(\mathbf{k}) = \Delta_{jj'} M_jC_{j'}$ only breaks $M_l$ ($l\neq j \neq j'$). Adding the $\Delta_j$ term modifies $h_j\mapsto h_j + \mathcal{c}_{j'}\mathcal{c}_l\Delta_j m_{jj}$ and results in a selective hinge gap closing along the $k_j$-hinge with chiral eigenvalues $\mathcal{c}_{j'}\mathcal{c}_l = \text{sgn}(\Delta_j)$. Similarly, adding $\Delta_{jj'}$ term modifies $h_j\mapsto h_j + \mathcal{c}_l\Delta_{jj'} m_{jj}$ resulting in a selective $k_j$-hinge gap closing with $\mathcal{c}_l = \text{sgn}(\Delta_{jj'})$. Thus, the selective nature of the hinge gap closing depends on the combination of the chiral eigenvalues $\mathcal{c_2c_3}$ and $\mathcal{c}_2$ for the $k_1$-hinge respectively.

In Fig.~\ref{fig:SEGC3D}, we show the energy spectrum and the probability density of near-zero-energy states with open boundary conditions and $L = 20$ sites in each direction and $f=-0.3$, for mirror-symmetry breaking terms $\Delta_1 M_1$ and $\Delta_{13} M_1C_3$. Both result in a $k_1$-hinge gap closing, leaving zero-energy bound states at corners with $\mathcal{c_2c_3}=-\sgn(\Delta_1)$ and $\mathcal{c}_2=-\sgn(\Delta_{13})$, respectively.

\section{Summary and outlook}\label{sec:sum}%
We have presented concrete models in the BDI class in which breaking anticommuting mirror symmetries, while preserving the chiral symmetry protecting their higher-order topology, yields gap closings in a subset of boundaries with the same orientation. These models include the prototypical quadrupole insulator~\cite{Benalcazar_2017a,Benalcazar_2017b,Seradjeh_2008b}, which has been realized experimentally in a variety of artificial platforms~\cite{Serra-Garcia_2018,Peterson_2018,Noh_2018,Imhof_2018,Fan_2019,Mittal_2019,Chen_2020,He_2020} and theoretically predicted in electronic materials
~\cite{Liu_2019,Lee_2020,Xue_2021,Guo_2022,Wang_2025}. We extended these results to nonseparable models with diagonal hopping elements and to other codimensions. Extensions to other classes and crystalline symmetries are interesting problems for future work.

The edge selection mechanism reported in this work can only operate in higher-order topologies by removing bound states from a subset of edges with codimension $>1$. For example, breaking the mirror symmetry $m_{jj}$ while preserving the chiral symmetry $c_j$ in $h_j$ results in bulk gap closing, which can only remove bound states in pairs from opposite edges. Similarly, the case of breaking (a subset of) diagonal mirror symmetries in our model, which commute with the chiral operator, results in bulk gap closing and cannot, by itself, result in edge selection.

We showed that the selective bulk-boundary correspondence can be understood within the low-energy theory of the model and that, moreover, it is furnished by an edge-sensitive topological invariant. Other topological invariants defined in the literature, such as the real-space invariant defined in Ref.~\cite{Benalcazar_2022} for chiral-symmetric models or Wilson-loop invariants based on mirror symmetries~\cite{Aich_2025b}, are not sensitive to edge orientations and cannot furnish the edge-selective bulk-boundary correspondence. Formulating an edge-sensitive Wilson-loop or real-space invariant is an interesting problem for future work.

\begin{acknowledgments}
We acknowledge fruitful discussions with T. L. Hughes, E. Khalaf, P. Zhu, and correspondence with W. Benalcazar, A. Cerjan, and L. Trifunovic in various stages of this work. This work is supported in part by Indiana University Office for Research Development (Bridge Program), IU Institute for Advanced Study, and by the National Science Foundation through Grant No. DMR-2533543.
\end{acknowledgments}

\appendix

\section{A winding number theorem}\label{app:winding}%
\label{app:winding}
If there is a symmetry $m$, $m h(k) m = h(\mathsf{m}k)$, $m^2 = 1$, that maps $k\in \ell \mapsto \mathsf{m}k\in \mathsf{m}\ell$ and commutes with the chiral operator, $mp_c m = p_c$, we have
\begin{align}
w_\ell[h,c] 
	&= \frac1{2\pi i} \oint_{\mathsf{m}\ell} \tr[p_c (m \underline h m) \partial_k (m\underline h m)] dk \nonumber \\
	&= \frac1{2\pi i} \oint_{\mathsf{m}\ell} \tr[(m p_c m) \underline h \partial_k \underline h] dk  \nonumber \\
	&= \frac1{2\pi i} \oint_{\mathsf{m}\ell} \tr(p_c \underline h \partial_k \underline h) dk  \nonumber \\
	&= w_{\mathsf{m}\ell}[h,c].
\end{align}
Thus, for a loop that is inverted under $m$, $\mathsf{m}\ell = - \ell$, we have $w_\ell[h,c] = w_{-\ell}[h,c] = -w_\ell[h,c] = 0$. 

Since $\ell'_j$ is inverted under $M'_{\bar j}$ and $[C,M'_{\bar j}] = 0$, the winding number $w_{\ell'_j}[H,C] = 0$. Since $\ell'_j$ is invariant under $M'_j$, we may use the diagonal basis of $M'_j$ to write $H\vert_{\ell'_j} = h'_{j+} \oplus h'_{j-}$ and  $C = c'_{j+}\oplus c'_{j-}$. This yields $w_{\ell'_j}[H,C] = \nu'_{j+} + \nu'_{j-}$, where $\nu'_{j\pm} = w_{\ell'_j}[h'_{j\pm},c'_{j\pm}]$ are the diagonal mirror winding numbers. Thus, $\nu'_{j+} + \nu'_{j-} = 0$.

\section{Spectrum with $\Delta'_1(M_1-M_2)$}\label{app:spec}%
\label{app:spec}
Since $M'_jM_lM'_j = (-1)^jM_{\bar l}$, we find that $M_1-M_2$ commutes with $M'_1$ and anticommutes with $M'_2$. Therefore, after adding this term to the $C_4$-symmetric Hamiltonian $H_0$, the Hamiltonian still commutes with $M'_1$ along the $(k,-k)$ diagonal, the loop $\ell'_1$ . We may, therefore, choose the diagonal basis of $M'_1$ along this diagonal to write $H\vert_{\ell'_1} = \sqrt2[\sigma_z\otimes h(k) + \Delta'_1 \id\otimes\sigma_x]$, which shows the energy eigenvalues along $(k,-k)$ are $\pm\sqrt2|d(k)\pm \Delta'_1|$.

Using the decomposition $h_j = m_{jj} d_{je} + im_{jj}c_j d_{jo}$, we find $[M_1,H_0(\vex k)] = i d_{1o}(k_1) M_1 (m_{11} c_1 \otimes \id)  = i d_{1o}(k_1) C$ and $[M_2,H_0(\vex k)] = i d_{2o}(k_2) M_2 (c_1\hspace{-0.5mm} \otimes m_{22} c_2) = i d_{2o}(k_2) C$. Thus, for the $C_4$-symmetric model along the $(k,k)$ diagonal, $[H\vert_{\ell'_2},M_1-M_2] = 0$. Therefore, the energy eigenvalues take the form $\pm\sqrt2(|d(k)|\pm\Delta'_1)$.

\bibliography{refs}

\clearpage

\end{document}